\documentclass[twocolumn,showpacs,preprintnumbers,amsmath,amssymb]{revtex4}
\usepackage{graphicx}
\usepackage{dcolumn}
\usepackage{bm}

\usepackage{epsfig}
\newcommand{\beq}{\begin{equation}}
\newcommand{\eeq}{\vspace{0cm} \end{equation}}
\newcommand{\beqq}{\setlength\arraycolsep{2pt}\begin{eqnarray}}
\newcommand{\eeqq}{\vspace{0cm} \end{eqnarray}}

\begin{document}

\title{Accelerating Cold Dark Matter Cosmology ($\Omega_{\Lambda}\equiv 0$)}

\author{J. A. S. Lima$^{1}$}\email{limajas@astro.iag.usp.br}

\author{F. E. Silva$^{2}$}\email{nalterii@ufrnet.br}

\author{R. C. Santos$^{1}$}\email{cliviars@astro.iag.usp.br}

\affiliation{$^{1}$Departamento de Astronomia, Universidade de S\~ao Paulo\\
Rua do Mat\~ao, 1226, 05508-900, S\~ao Paulo, SP, Brazil}

\affiliation{$^{2}$Departamento de F\'{\i}sica Te\'orica e Experimental (UFRN)\\
C. P. 1641, 59072-970, Natal, RN, Brazil}

\begin{abstract}\begin{center}{\bf Abstract}\end{center}
A new kind of accelerating flat model with no dark energy
that is fully dominated by cold dark matter (CDM) is investigated. 
The number of CDM particles is not conserved 
and the present accelerating stage is a consequence of 
the negative pressure describing the irreversible process of gravitational particle creation. 
A related work involving  accelerating CDM cosmology  has been  discussed before the SNe observations 
[Lima, Abramo \& Germano, Phys. Rev. D53, 4287 (1996)]. However, in order to have a transition from a decelerating to an 
accelerating regime at low redshifts, the matter creation rate proposed here includes a constant term  
of the order of the Hubble parameter.  In this case, $H_0$ does not need to be small in order to solve the age problem 
and the transition happens even if the matter creation is negligible during the radiation and part of the matter dominated phase. Therefore, instead of the vacuum dominance at redshifts 
of the order of a few, the present accelerating 
stage in this sort of Einstein-de Sitter CDM cosmology is a consequence of the  
gravitational particle creation process.  As an extra bonus, in the present scenario does not exist the 
coincidence problem that plagues models with dominance of dark energy.  The model is able to harmonize
a CDM picture with  the present age of the universe,  the latest measurements of the Hubble parameter and the Supernovae observations.

\end{abstract}

\pacs{98.80.-k, 95.35.+d,95.30.Tg}

\maketitle

\section{Introduction}

A large amount of data relevant to cosmology (involving Supernovae type Ia and cosmic 
background radiation probes) have provided strong evidence that the observed universe is 
well described by an accelerating,  flat Friedmann-Robertson-Walker (FRW) model \cite{Riess,Riess07,CMB,Ages}. 
However, the substance or mechanism behind the current cosmic acceleration remains unknown and constitutes a challenging problem of modern cosmology. 

In relativistic cosmology, an accelerating regime is obtained 
by assuming the existence of a dark energy component 
(in addition to cold dark matter), an exotic fluid endowed with negative pressure in order to violate 
the strong energy condition \cite{review}. The simplest theoretical
representation of dark energy is by means of a cosmological constant $\Lambda$,
which acts on the Einstein field equations (EFE) as an isotropic and
homogeneous source with constant equation of state (EoS) $w
\equiv p/\rho = -1$. 

All observational data available so far seems to be in 
good agreement with the  cosmic concordance model, i.e.,  
a vacuum energy plus cold dark matter ($\Lambda$CDM) scenario. 
Nevertheless, $\Lambda$CDM models are plagued with several 
problems. For instance, it is very
difficult to reconcile the small value required by observations
($\simeq 10^{-10} \rm{erg/cm^{3}}$) with estimates from quantum
field theories ranging from 50-120 orders of magnitude larger
\cite{weinberg}.   Such problem has inspired many authors to propose  
alternative candidates in the literature \cite{Sfield,decaying, decaying1,list1,XM1}, among them: (i) a relic scalar field slowly rolling down its potential, (ii) a $\Lambda
(t)$-term or a decaying vacuum energy density, (iii) the
``X-matter'',  an extra component characterized by equation
of state $p_x=\omega \rho_x$, where $\omega$ may be constant or a redshift dependent function, (iv) a Chaplygin-type
gas whose equation of state is $p=-A/\rho^{\alpha}$, where 
A and $\alpha$ are positive parameters.  More recently, some attention has also 
been paid to a possible interaction between  the dark sector components \cite{DSI}. 

The space parameter of such  models are usually highly degenerated and some of 
them contain the $\Lambda$CDM scenario as a particular case. 
In point of fact, the plethora of possible candidates does  not help to identify  the 
nature of this mysterious component since there is no compelling  direct evidence yet 
for dark energy (or its dynamical effects). In other words, 
the evidence supporting its existence is not strong enough to be considered 
established beyond doubt (see \cite{BLB} for a critical discussion).  

Roughly speaking, a realistic cosmological scenario should be in agreement with at 
least four well established observational results, namely: (i) the existence 
of a dark non-baryonic component as required by the dynamics of galaxies and clusters, 
the matter power spectrum and other independent probes like the temperature anisotropies 
of the cosmic microwave background (CMB) from last scattering surface (ii) the late time cosmic acceleration, 
(iii) the (nearly) flatness of the Universe, and, finally, 
(iv) a Hubble parameter $H_0 \approx 72\;\mbox{km/s.Mpc}$ with 
the Universe being older than 12 Gyrs in order to accommodate the 
oldest observed structures (globular clusters). When confronted with this simple requirements, we see that the CDM 
or Einstein-de Sitter cosmology is in  clear contradiction with results (ii) and (iv).  Therefore, if one assumes that 
the dark energy does not exist, the first task is to explain how a 
flat CDM dominated Universe can accelerate at late times because, potentially,  
accelerating cosmologies solve the age problem. 

In this concern, we recall that the presence of a negative pressure is the key ingredient required to 
accelerate the expansion. This kind of stress  occurs naturally in many different contexts when  the physical systems 
depart from a thermodynamic equilibrium states \cite{LL}.   
In general, such states are connected with phase 
transitions (for example, in an overheated van der Waals liquid), and for some systems 
the existence of negative pressure seems to be inevitable \cite{sakharov}. 
In this connection, as first pointed out by Zeldovich \cite{zeld}, the 
process of  cosmological particle  creation at the expenses of the gravitational field can phenomenologically be 
described by a  negative pressure and the associated entropy production. 
In principle, such an approach is completely different from the 
one developed  by  Hoyle and Narlikar \cite{narlikar} adding extra terms to the Einstein-Hilbert action 
describing the so-called C-field.  In the latter case, the creation 
phenomenon is explained trough  a process of interchange of energy and 
momentum between matter itself and the C-field as happens, for instance, in 
vacuum decaying cosmologies \cite{decaying}. 

The gravitational matter creation processes was 
investigated from a microscopic viewpoint by Parker and collaborators \cite{Parker} by considering
the Bogoliubov mode-mixing technique in the context of quantum field theory 
in curved space-time \cite{BirrellD}. Despite being rigorous and well-motivated, 
those models were never fully realized,  probably due to 
the lack of a well-defined  prescription of how matter creation is to be incorporated 
in the classical EFE. 
  
The consequences of gravitational matter creation have also been macroscopically 
investigated mainly as a byproduct of bulk viscosity processes near the Planck era 
as well as during the reheating of  the  inflationary scenarios \cite{Murphy,Hu}.     
However, the first self-consistent macroscopic formulation of the matter creation process 
was  put forward by Prigogine and coworkers \cite{Prigogine} and somewhat 
clarified by Calv\~{a}o, Lima and Waga \cite{LCW} through a manifestly covariant formulation. 
It was also shown that matter creation, at the expenses of the gravitational field,   
can effectively be discussed in the realm of the relativistic nonequilibrium thermodynamics. Later on, it was 
also demonstrated that the matter creation is an irreversible process 
completely  different from bulk viscosity mechanism \cite{LG92} (see also \cite{SLC02} for a more complete discussion).
Several interesting  features of cosmologies with creation of matter and radiation have been investigated by many authors \cite{ZP2,LGA96,LA99,Susmann94,ZSBP01,freaza02} (see also \cite{Makler07} for recent studies on this subject).

In comparison to the standard equilibrium 
equations, the irreversible creation process is 
described by two new ingredients: a balance 
equation for the particle 
number density and a negative  
pressure term in the stress tensor. Such quantities 
are related to each other in a very definite way by the second law of 
thermodynamics \cite{Prigogine,LCW}.  The {\it leitmotiv} of this
approach is that the matter creation process, at the expense of the
gravitational field, can happen only as an irreversible process
constrained by the usual requirements of non-equilibrium
thermodynamics. 

In this context, we are proposing here a new flat cosmological scenario where the cosmic acceleration is 
powered uniquely by the  creation of cold dark matter particles. As we shall see, the model 
is consistent with the supernovae type Ia data, and a transition redshift 
of the order of a few is also obtained. In this extended CDM model, 
the Hubble parameter does not need to be small in order to solve the age problem  
and the transition happens even if the matter creation is negligible during the 
radiation and considerable part of the matter dominated phase. 
Moreover, the so-called coincidence problem of dark energy models 
is replaced here by a  gravitational particle creation process at 
low redshifts.

\section{Cosmology and Matter Creation}

For the sake of generality, let us  start with the homogeneous and isotropic FRW line element
\begin{equation}
\label{line_elem}
  ds^2 = dt^2 - R^{2}(t) (\frac{dr^2}{1-k r^2} + r^2 d\theta^2+
      r^2sin^{2}\theta d \phi^2),
\end{equation}
where $R$ is the scale factor and $k= 0, \pm 1$ is the curvature 
parameter. Throughout we use units such that $c=1$.

In that background, the nontrivial EFE for a fluid endowed with 
matter creation and the balance equation for the particle number 
density can be written as \cite{Prigogine,LCW,LG92,ZP2}

\begin{equation}
    8\pi G \rho = 3 \frac{\dot{R}^2}{R^2} + 3 \frac{k}{R^2},
\end{equation}

\begin{equation}
   8\pi G (p+p_{c}) = -2 \frac{\ddot{R}}{R} - \frac{\dot{R}^2}{R^2} -
	\frac{k}{R^2},
\end{equation}

\begin{equation}\label{number}
      \frac{\dot{n}}{n} + 3 \frac{\dot{R}}{R} = 
           \frac{\psi}{n}\equiv \Gamma,
\end{equation}
where an overdot means time derivative and $\rho$, $p$, $p_{c}$, $n$ and
$\psi$ are the energy density, thermostatic pressure, creation pressure,
particle number density and matter creation rate, respectively. The quantity $\Gamma$ with dimension of 
$(time)^{-1}$ is the creation rate of the process. The creation pressure 
$p_{c}$ is defined in terms of the creation rate and other physical 
quantities.  In the case of adiabatic 
matter creation, it is given by \cite{Prigogine,LCW,LG92,ZP2,SLC02} (see also Appendix A for a simplified deduction)
 
\begin{equation}\label{CP}
    p_{c} = - \frac{\rho + p}{3nH} \psi\equiv -\frac{\rho + p}{3H} \Gamma,
\end{equation}
where $H = {\dot {R}}/R$ is the Hubble parameter.

As one may check, by combining the EFE with usual equation of state, $p = \omega \rho$, 
the equation governing the evolution 
of the scale function is readily obtained:

\begin{equation}
\label{evolR}
     R{\ddot R}+ \left[\frac{1 + 3\omega}{2} - \frac{(1 + \omega)
     \Gamma}{2H}\right]\left(\dot{R}^2 + k \right) = 0.  
\end{equation}
The above expression shows how the matter creation rate,  $\Gamma$, modifies the 
evolution of the scale factor as compared to the case with no creation. Conversely,  
the cosmological dynamics with irreversible matter creation will be defined once the matter creation rate is given. 
As should be expected, by taking  $\Gamma=0$ it reduces to the FRW differential equation governing the evolution of a perfect simple fluid \cite{LA}.

\section {Flat CDM model with matter creation and the age of the Universe}

In what follows we focus our attention on the flat cold dark matter model ($k=\omega=0$) with the previous equation reducing to: 
\begin{equation}
\label{evolR}
     R\ddot{R} + \frac{1}{2}\left(1 - \frac{\Gamma}{H}\right) \dot{R}^2 = 0,  
\end{equation}
or, equivalently,
\begin{eqnarray}\label{H}
\dot{H}+\frac{3}{2}H^2 \left(1-\frac{\Gamma}{3H}\right)=0.
\end{eqnarray}

On the other hand,  Eq. ($\ref{number}$) can be rewritten as 

\begin{equation}
{\dot n \over 3\,n\,H}  \ + \ 1 \ = \ {\Gamma \over 3H},
\label{w23}
\end{equation}
which means that the creation process can effectively be quantified by the dimensionless ratio (see also Eq. (\ref{H}))
\begin{equation}
\Delta (t) = \frac{\Gamma}{3H},
\end{equation} 
which in general is a function of time. If  $\Gamma \ll 3H$, that is,  $\Delta \ll 1$, the  creation process is negligible leading 
to  $n \propto R^{-3}$ and $H=2/3t$,  as should be expected for an Einstein-de Sitter model.  
The opposite regime ($\Gamma \gg 3H$) defines an extreme
theoretical situation, where the creation process is a  phenomenon so powerful that
the dilution due to expansion is more than compensated. Probably, this kind of behavior may happen 
only in the very early universe as happens, for instance, during the  reheating stage of inflation. 
An intermediary (and physically more reasonable situation) occurs if this ratio is smaller or of the order
of unity ($\Gamma \lesssim  3H$). In particular, if $\Gamma=3H$ the dilution due to expansion is exactly compensated 
and the number density remains constant. From now on we consider that $\Delta(t)\leq 1$.  

\begin{figure}[!ht]
\begin{center}
\epsfig{file=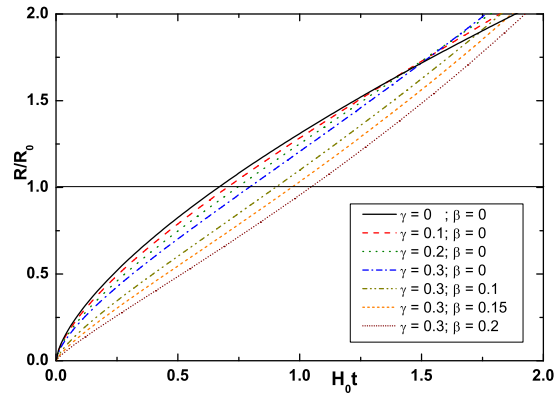, height=6.0cm, width=8.0cm}
\caption{The scale function as a function of time. Like the FRW dust filled model (solid black line), 
the evolution starts from an initial  singularity. Note that the influence of $\gamma$ is 
negligible in the early universe (see Eq. (\ref{H1}) for $H \gg H_0$).} \label{figure1}
\end{center}
\end{figure}

In a series of papers \cite{LGA96,LA99}, we have investigated some properties of adiabatic matter creation models with $\Gamma=3\beta H$,  
where $\beta$ is a constant parameter contained on the interval [0,1] ($\Delta=\beta$).  However, that kind of models are 
always accelerating for $\beta > 1/3$ or decelerating for 
$\beta < 1/3$, that is, there is not a transition redshift from a decelerating to an accelerating regime as required 
by the supernovae type Ia observations (see Figure 3a).  
In order to cure such a  difficulty  we add a constant term in this expression, that is, we consider the 
following matter creation rate (see Appendix for a more rigorous argument) 
\beq 
{\Gamma}=3\gamma H_0+3\beta H \label{Gamma}, 
\eeq 
where the parameter $\gamma$ (like $\beta$) lies on the interval [0,1]. As we shall see, 
this scenario is compatible with the basic observations listed in the introduction even for $\beta=0$. 
Inserting Eq. (\ref{Gamma}) into (\ref{H}) one finds  
\begin{eqnarray}
\dot{H} + \frac{3}{2}H^2\left(1-\beta-\frac{\gamma
H_0}{H}\right)=0,\label{H1}
\end{eqnarray}
whose solution reads
\beq
H(t)=H_0\left(\frac{\gamma}{1-\beta}\right)\frac{e^{\frac{3\gamma
H_0}{2}t}}{(e^{\frac{3\gamma H_0}{2}t}-1)},
\label{H2} 
\eeq 
and by integrating  the above expression  we obtain a big-bang solution for the scale factor
\beq R(t)=R_0\left[
\left(\frac{1-\gamma-\beta}{\gamma}\right)(e^{\frac{3\gamma
H_0}{2}t}-1)\right]^{\frac{2}{3(1-\beta)}},\label{R} 
\eeq 
where $R_0$ and $H_0$ are the present day values of $R(t)$ and $H(t)$, respectively.  
In the limit  $\gamma\rightarrow 0$, the above expression reduces to
\beq
R(t)=R_0\left[\frac{3}{2}(1-\beta)H_0t\right]^{\frac{2}{3(1-\beta)}},\label{R1}
\eeq 
which is the model discussed in Refs. \cite{LGA96,LA99}, and as should be expected the Einstein-de Sitter cosmology is recovered for $\beta=0$. 

\begin{figure}[!ht]
\begin{center}
\epsfig{file=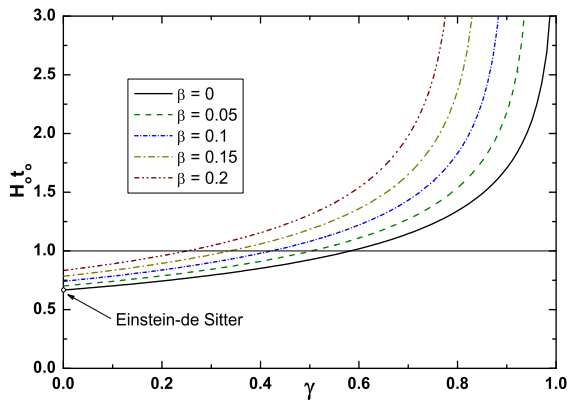, height=6.0cm, width=8.0cm}
\caption{The age of the Universe as a function of the $\gamma$ parameter and some fixed values of $\beta$. Note that ages great enough are  obtained even for $\beta=0$ (bottom solid line). For a given value of $\gamma$, the effect of the $\beta$ parameter is to increase the age of the Universe.} \label{figure1}
\end{center}
\end{figure}
In Figure 1 we display the behavior of the scale factor as a function of time. All the models start their evolution from the initial singularity ($R(0)=0$). It is worth noticing that the $\gamma$ parameter does not contribute at early times. Actually, for $H \gg H_0$ only the $\beta$ parameter appears in the equation of motion (\ref{H1}).  

Now, by taking $H=H_0$ in Eq. (\ref{H2}) or $R=R_0$ in (\ref{R}), the following expression for the age of the Universe is readily obtained
\begin{eqnarray}
t_0=H_0^{-1}\frac{2}{3\gamma}\ln\left(\frac{1-\beta}{1-\gamma-\beta}\right).
\label{age}
\end{eqnarray}
which for $\gamma=0$ reduces to $H_0t_0=2/3(1 -\beta)$ as expected (see \cite{LGA96,LA99}).

In Figure 2 we show the age parameter as a function of $\gamma$ and some particular values of $\beta$. 
The solid black line  yields the age of the Universe as a function of $\gamma$ when $\beta$ is zero. In this case, 
\begin{eqnarray}
H_0t_0=\frac{2}{3\gamma}\ln\left(\frac{1}{1-\gamma}\right).
\label{age1}
\end{eqnarray}
Note that ages great enough are obtained even for $\beta=0$. In particular, for $\gamma = 0.6$ the age parameter is $H_0t_0 =1$, 
exactly the same value predicted by the `cosmic concordance' ($\Lambda$CDM) model from WMAP3 and complementary observations \cite{CMB}.  In the limit $\gamma \rightarrow 0$ one  obtains $H_0t_0 =2/3$ as should be expected. The influence of the $\beta$ parameter is apparent from Figure 2, namely, it increases the age of the Universe for a given value of $\gamma$. 

At this point, it is interesting to discuss in what sense this simple CDM scenario with creation behaves like an irreversible process. 
Adiabatic matter creation means that the total entropy S increases, but, the specific entropy (per particle), $\sigma =
S/N$, where $N$ is the corresponding number of particles, remains constant \cite{Prigogine,LCW}. 
Quantitatively, $\dot\sigma =0$ implies that 

\begin{equation}\label{S0}
\frac{\dot S}{S} = \frac{\dot N}{N}.
\end{equation}
Hence, due to the creation processes ($\dot N>0$), the universe does not expand
adiabatically as happens in the standard
CDM model. Besides, since
up to a constant factor one has $N=nR^3$, by inserting Eq. (\ref{Gamma}) into (\ref{number}) a straightforward integration yields

\begin{equation}
N(t)=N_o {( \frac{R}{R_o} )}^{3\beta}e^{3\gamma H_0 (t -t_0)}.
\end{equation}

Further, from Eq. (\ref{S0}), $S=S_o (N/N_o)$, and using the above
expression one may write the entropy of the CDM particles like 

\begin{equation}\label{S}
S(t)= S_{0}{( \frac{R}{R_o} )}^{3\beta}e^{3\gamma H_0 (t -t_0)},
\end{equation}
where $S_{0}$ is the present entropy of the CDM fluid. Note that if $\gamma=\beta=0$ the standard conserved quantities are recovered. 

\begin{figure*}
\centerline{\psfig{figure=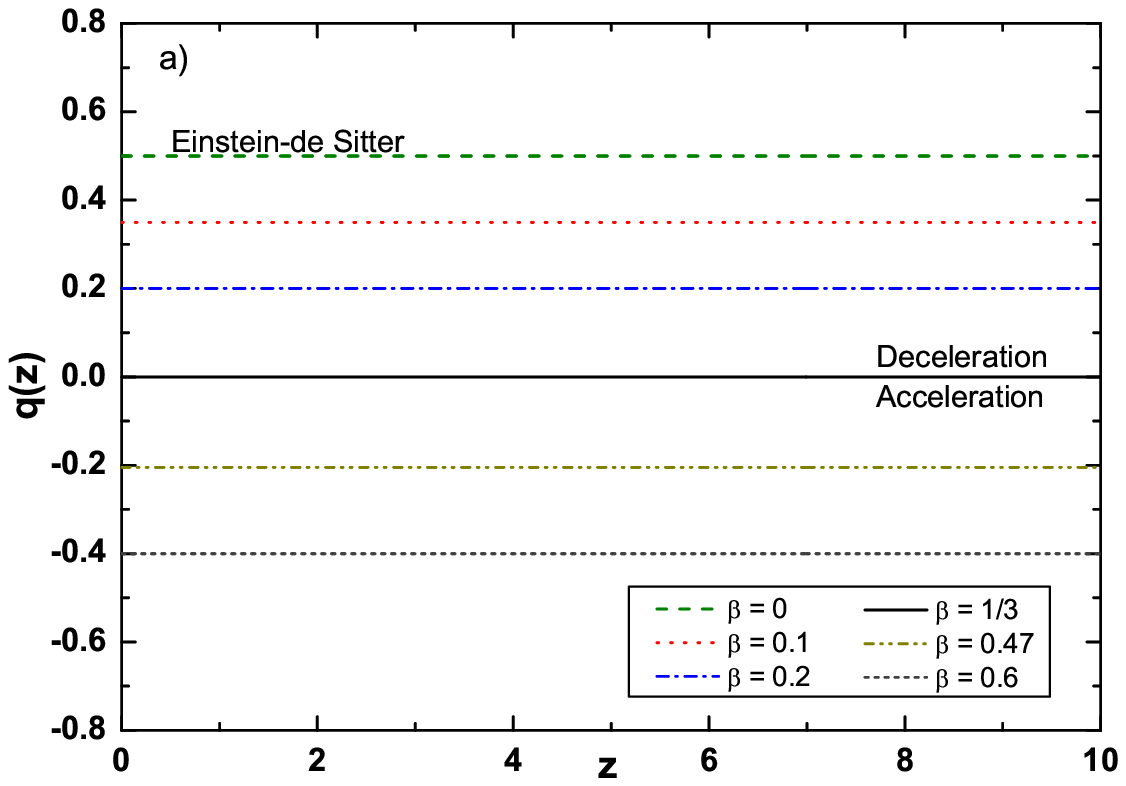,width=3.5truein,height=2.4truein}
\psfig{figure=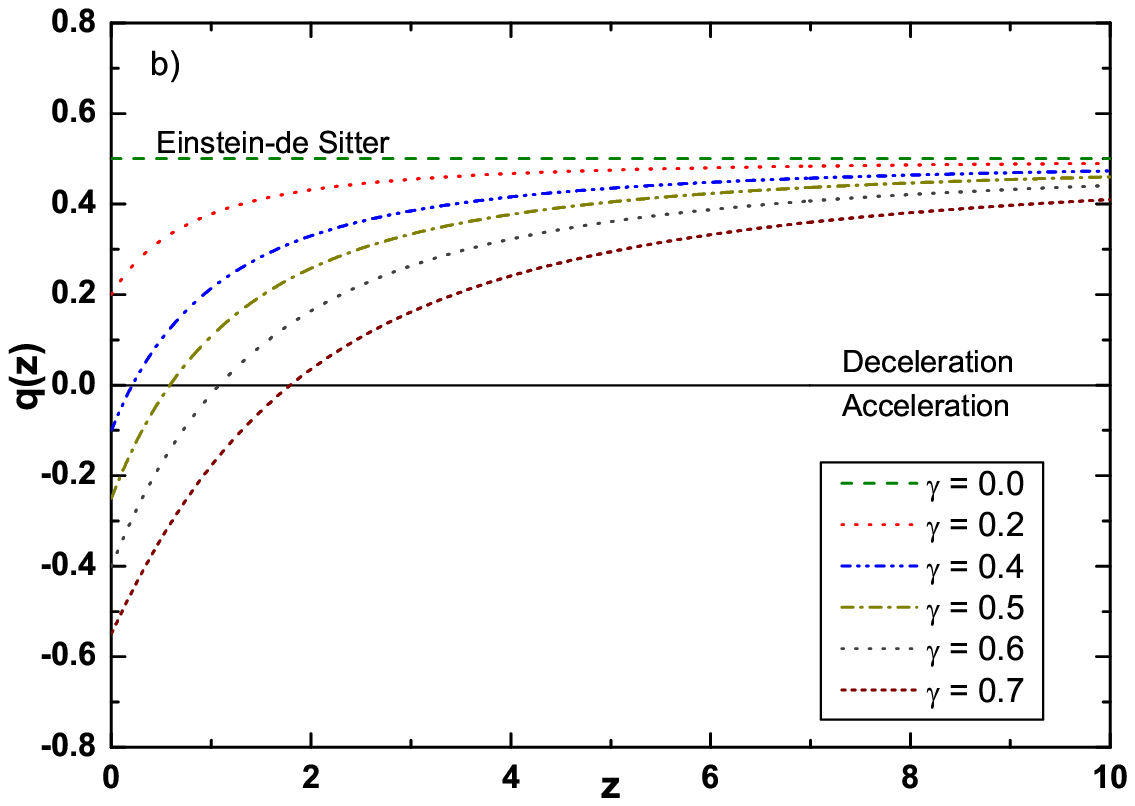,width=3.5truein,height=2.4truein}\hskip
0.1in} \caption{{\bf{a)}} Effect of the $\beta$ parameter on q(z). For all curves $\gamma$ was taken to be zero (see Eq. (\ref{q2})). Note that $\beta = 1/3$ is a critical value for which $q=0$. For $\beta < 1/3$  the possible values of q(z) are always constant and positives while for $\beta > 1/3$ they remain constant and negatives in the course of the expansion.  There is no transition redshift in this case.  {\bf{b)}} Effect of the $\gamma$ parameter on q(z). In this case, the $\beta$ parameter has been taken to be zero. The creation of CDM particles is negligible at high redshifts. Due to the particle creation at redshifts of the order of a few occurs a transition from a decelerating to an accelerating regime.}\label{fig3}
\end{figure*}

\section{Decelerating Parameter, Transition Redshift and Supernova Bounds}

To begin with, we first observe that by combining Eqs. (\ref{evolR}) and (\ref{Gamma}), the decelerating parameter  reads 
\beq 
\label{q1} 
q=\frac{1}{2}\left[1-3\beta - 3\gamma\frac{H_0}{H}\right],
\eeq
so that for $\gamma=0$ the value of $q$ remains constant as remarked earlier. Now, by eliminating the time from Eqs. (\ref{H2}) and (\ref{R}), and using that 
$R=R_0 (1+z)^{-1}$, one obtains the Hubble parameter in terms of the redshift 
\beq 
H(z)=H_0\left[\frac{\gamma
+(1-\gamma-\beta)(1+z)^{\frac{3}{2}(1-\beta)}}{1-\beta}\right],\label{Hz}
\eeq 
and inserting this result into (\ref{q1}) it follows that
\beq 
q(z)=\frac{1}{2}
\left[\frac{(1-3\beta)(1-\gamma-\beta)(1+z)^{\frac{3}{2}(1-\beta)}-2\gamma}{(1-\gamma-\beta)(1+z)^{\frac{3}{2}(1-\beta)}+\gamma}\right].\label{q2}
\eeq
For $\gamma=0$, this expression yields  $q= (1-3\beta)/2$, while for $\beta=0$ we find
\beq q(z)=\frac{1}{2}
\left[\frac{(1-\gamma)(1+z)^{\frac{3}{2}}-2\gamma}{(1-\gamma)(1+z)^{\frac{3}{2}}+\gamma}\right].\label{q3}
\eeq
In Figure 3 we display the decelerating parameter as a function of the  redshift as given by  the above expressions. As remarked earlier,  the existence of a transition redshift at late times depends exclusively on the $\gamma$ parameter (compare Figs. 3a and 3b). 

A simple relation uniting $\gamma$, $\beta$ and $z_t$ can be determined by taking $q=0$. As one may check, Eq. (\ref{q2}) implies that
\beq
z_t=\left[\frac{2\gamma}{(1-3\beta)(1-\gamma-\beta)}\right]^{\frac{2}{3}(1-\beta)}-1,\label{zt}
\eeq
or equivalently,
\beq 
\gamma=\frac{(1-3\beta)(1-\beta)(1+z_t)^{\frac{3}{2}(1-\beta)}}{2+(1-3\beta)(1+z_t)^{\frac{3}{2}(1-\beta)}}. \label{gamma}
\eeq 
For $\beta=0$ the above expression reduces to 
\beq \label{gammazt}
\gamma=\frac{(1+z_t)^{\frac{3}{2}}}{2+(1+z_t)^{\frac{3}{2}}}, 
\eeq
and the age of the Universe can be rewritten in terms of the transition redshift. One finds,
\begin{eqnarray}
t_0=H_0^{-1}\frac{4+2(1+z_t)^{\frac{3}{2}}}{3(1+z_t)^{\frac{3}{2}}}\ln\left[1+\frac{(1+z_t)^{\frac{3}{2}}}{2}\right].
\end{eqnarray}

\subsection{Constraints from SNe Ia Observations}
Let us now discuss the constraints from distant type Ia SNe data  on the class of CDM accelerating cosmologies proposed here. 
Since $H_0$ can be determined from the Hubble Law and $\Omega_M=1$, the model has only two
independent parameters, namely, $\gamma$ and $\beta$ (see Eq. (\ref{Hz}) for $H(z)$). 

The predicted distance modulus for a supernova at redshift $z$, given a set of
parameters $\mathbf{s}$, is
\begin{equation} \label{dm}
\mu_p(z|\mathbf{s}) = m - M = 5\,\mbox{log} d_L + 25,
\end{equation}
where $m$ and $M$ are, respectively, the apparent and  absolute
magnitudes, the complete set of parameters is $\mathbf{s} \equiv
(H_0, \gamma, \beta)$, and $d_L$ stands for the luminosity distance (in
units of megaparsecs),
\begin{equation}
d_L = c(1 + z)\int_{x'}^{1} {dx
\over x^{2}{H}(x;\mathbf{s})},
\end{equation}
with $x' = {R(t) \over R_0} = (1 + z)^{-1}$ being a  convenient
integration variable, and ${H}(x; \mathbf{s})$ the expression given
by Eq. (\ref{Hz}).

\begin{figure*}
\centerline{\psfig{file=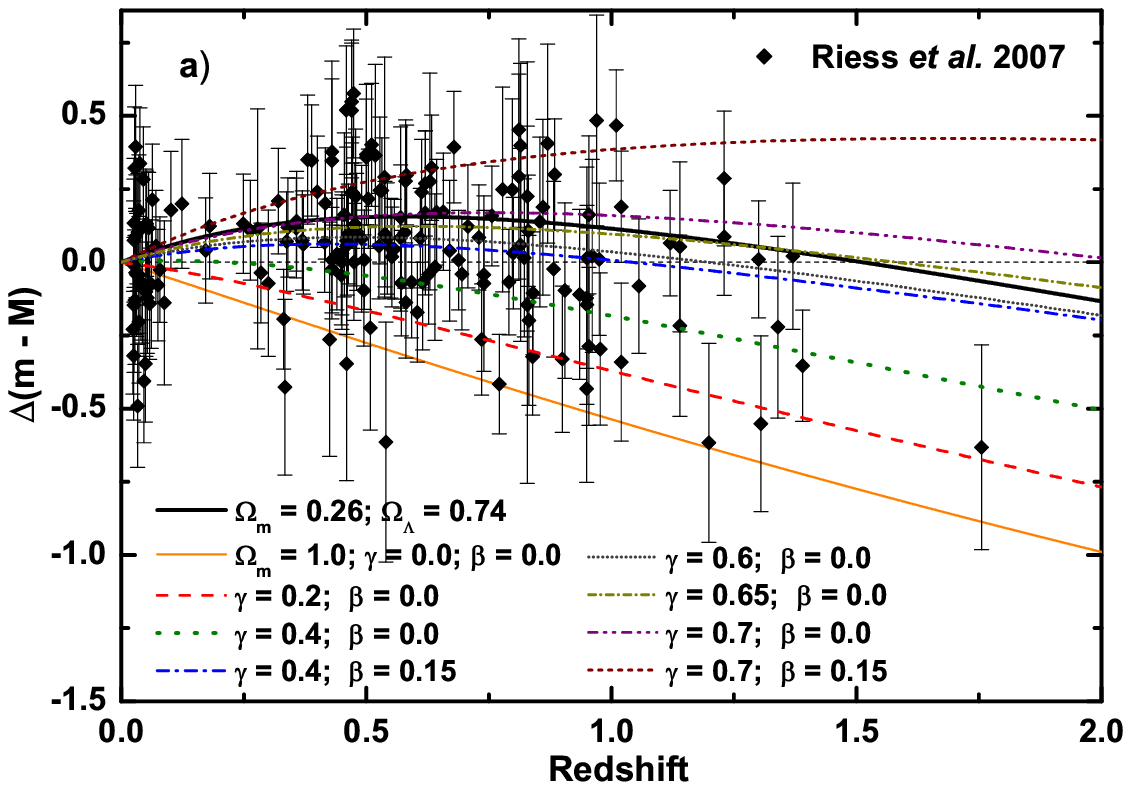, width=3.5truein, height=2.4truein}
\psfig{figure=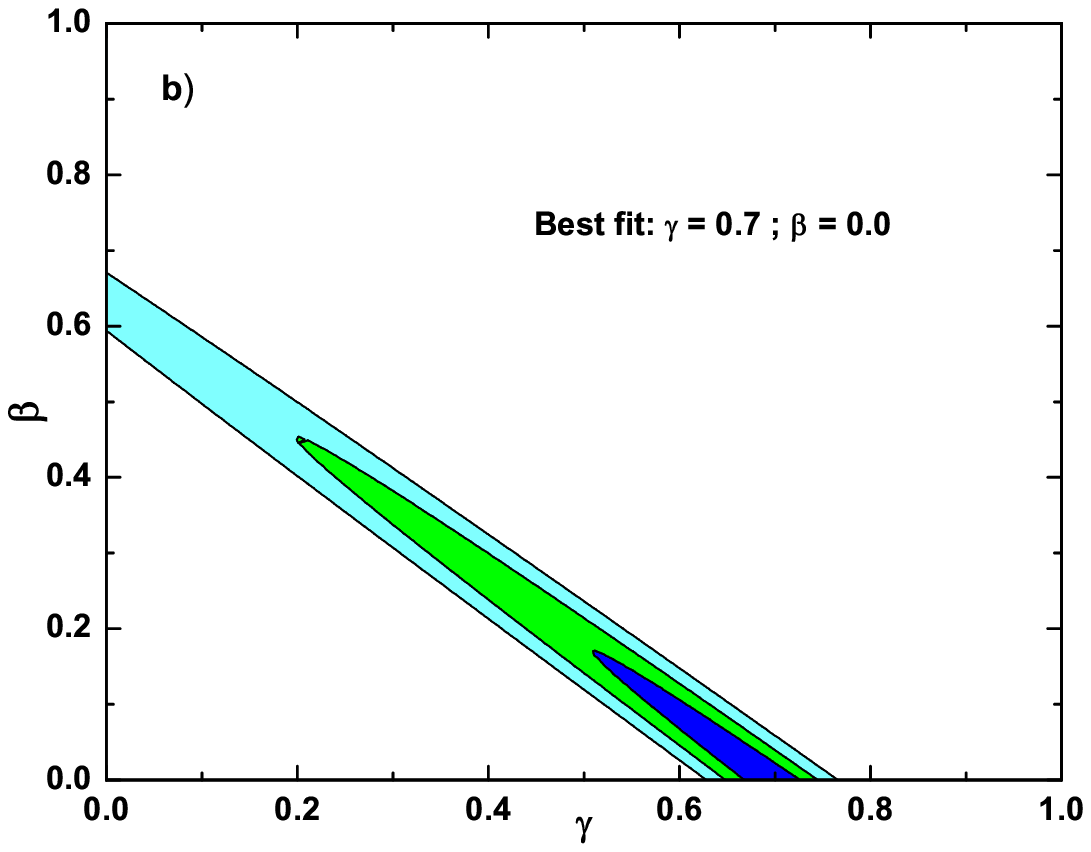, width=3.5truein, height=2.4truein}\hskip
0.1in} \caption{{\bf {a)}} Residual magnitudes (relative to an empty model) of 182 Supernovae data from Riess et al. (2007)  and the predictions of the accelerating CDM model for several values of the pair ($\gamma$,$\beta$). For comparison we display the Einstein-de Sitter (bottom curve) and the $\Lambda$CDM (solid black curve). Note that the curve for $\gamma =0.65$ and $\beta=0$ (only one free parameter) is very close to the one of cosmic concordance ($\Lambda$CDM) scenario.  {\bf{b)}} The $\gamma$-$\beta$ plane for a flat CDM model with gravitational particle creation obtained from the same sample.  It should be stressed that the Supernova data can be fitted with just one free parameter ($\gamma$) which is responsible for the transition at late times (see Fig. 3b and comments on the main text and Appendix B).} \label{figure4}
\end{figure*}

In order to constrain the free parameters of the model consider now the latest sample 
containing 182 Supernovas as published by  Riess and coworkers  \cite{Riess07}.
The best fit to the set of parameters $\mathbf{s}$ can be estimated by
using a $\chi^{2}$ statistics with
\begin{equation}
\chi^{2} = \sum_{i=1}^{N}{\frac{\left[\mu_p^{i}(z|\mathbf{s}) -
\mu_o^{i}(z)\right]^{2}}{\sigma_i^{2}}},
\end{equation}
where $\mu_p^{i}(z|\mathbf{s})$ is given by Eq. (\ref{dm}),
$\mu_o^{i}(z)$ is the extinction corrected distance modulus for a
given SNe Ia at $z_i$, and $\sigma_i$ is the uncertainty in the
individual distance moduli.  By marginalizing on the nuissance parameter $h$ ($H_0 = 100h Km.s^{-1}.Mpc^{-1}$) we find $0.21 \leq \gamma \leq 0.75$ and $0 \leq \beta \leq 0.46$ at
$95\%$ of confidence level. The best fit adjustment occurs for values of $\gamma=0.7$ and 
$\beta=0$ with $\chi^{2}_{min}=175.8$ and $\nu=180$ degrees of freedom. The reduced $\chi^{2}_{r}=0.98$ where ($\chi^{2}_{r}=\chi^{2}_{min} /\nu$), thereby showing that the model provides a very good fit to these data.

\section{Conclusion}

In this paper we have proposed a flat cold dark matter cosmology whose late time acceleration is 
powered by an irreversible creation of CDM particles. In our scenario there is no dark energy, and, as such, the so-called coincidence problem is also absent.  
It should be stressed that $H_0$ does not need to be small in order to solve the age problem. Further, the transition from a decelerating to an accelerating regime at late times happens even if the matter creation is negligible during the radiation and considerable part of the matter dominated phase (this is equivalent to take $\beta = 0$ in all the expressions). Therefore, like in flat $\Lambda$CDM scenarios, there is just one free parameter, and the resulting model provides an excellent fit to the observed dimming of distant type Ia supernovae (see Figs. 4a and 4b).  Note also that the flat model ($\Omega_m=1$) with creation of CDM particles proposed here can easily be extended  to include negative ($\Omega_m <1$) and positive ($\Omega_m >1$) spatial curvatures. The same happens with the inclusion of a small (conserved) baryonic component whose density parameter today is severely constrained by the primordial nucleosynthesis and WMAP results. In this case, the value of the transition redshift as derived in section IV will be slightly modified. 

On other hand, the existence of such a model also means that the accelerating expansion does not represent a direct evidence 
for a non-zero cosmological constant or, 
more generally, to the existence of dark energy as usually assumed by many authors. 
Naturally, new constraints on the relevant parameters ($\gamma$ and $\beta$) from complementary observations need to be investigated  
in order to see whether the matter creation model proposed here provides a realistic description of the observed Universe.   
New bounds on these parameters coming from the background and perturbed equations in the presence of a conserved baryonic 
component will be discussed in a forthcoming communication. 

\appendix

\section{Particle Creation and Irreversibility} 

{In this appendix we describe how the creation pressure given by Eq. (\ref{CP}) can be deduced by using the 
relativistic non-equilibrium thermodynamics. The idea is to show in a simplified way how an irreversible mechanism of quantum origin can be 
incorporated in the classical Einstein field  equations. 

A relativistic self-gravitating  simple fluid endowed only with gravitational matter creation is characterized by an energy momentum
tensor $T^{\alpha \beta}$, a particle current $N^{\alpha }$, and an
entropy current $S^{\alpha}$. In the homogeneous and isotropic case, these quantities satisfy the following relations:
\begin{equation}
T^{\alpha \beta}=(\rho + p + p_c)u^{\alpha} u^{\beta} - pg^{\alpha \beta},
\quad T^{\alpha \beta};_{\beta}=0 , \label{eq:TAB}
\end{equation}
\begin{equation} \label{eq:NA}
N^{\alpha}=nu^{\alpha}, \quad  N^{\alpha};_{\alpha}= n\Gamma ,
\end{equation}
\begin{equation} \label{eq:SA}
S^{\alpha}=n\sigma u^{\alpha}, \quad  S^{\alpha};_{\alpha}= \tau \geq 0 ,
\end{equation}
where ($;$) means covariant derivative, $p_c$ is the creation pressure, $n$ is the particle number
density, $\Gamma$ is the particle creation rate (from quantum gravitational origin) $\sigma$ is the specific entropy (per particle), and $\tau$ is the entropy source. 
In what follows it is assumed that the particles
spring up into space-time in such a way that they
turn out to be in thermal equilibrium with the already
existing ones. The entropy production is then
due only to the scalar process of  matter creation (bulk viscosity has been neglected). Naturally, 
for  $\Gamma=0$ we shall expect that the creation pressure
vanishes and so also the entropy production. 

In the FRW background,  conditions  (\ref{eq:TAB}) and (\ref{eq:NA}) can be written as (a dot means comoving
time derivative)

 \begin{equation}\label{BE}
 \dot{\rho} + 3H (\rho + p + p_c)=0,\,\,\,\,\,
 \dot{n} + 3Hn=n\Gamma.
\end{equation}
The basic aim here is to show how the second law of thermodynamics constrains the dependence of $p_c$ on $\Gamma$ and other quantities specifying the fluid.  
Following standard lines, the quantities $p$, $\rho$, $n$ and $\sigma$  are related to the
temperature $T$ by the Gibbs law 

\begin{equation} \label{eq:GIBBS}
nTd\sigma=d\rho - {\rho + p \over n}dn,
\end{equation} 
while the chemical potential is defined by the Euler's relation 

\begin{equation} \label{eq:17}
\mu= {\rho+p\over n}-  T\sigma. 
\end{equation}

Now, by using equations (\ref{eq:SA})-(\ref{eq:17}) it is easy to show that the source of entropy reads   
\begin{equation}
\tau\equiv n\sigma \Gamma + n\dot\sigma=-\frac{3Hp_c}{T} - \frac{\mu n \Gamma}{T}\geq 0,
\end{equation}

Finally, the case of adiabatic gravitational matter creation means that the entropy increases but the specific entropy $\sigma$ remains constant ($\dot \sigma=0$). Therefore, the above equation  implies that  $\tau=n\sigma \Gamma \geq 0$ with the creation pressure assuming the form adopted in the present work (cf. Eq. (\ref{CP}))
\begin{equation}\label{CP1}
    p_{c} =  -\frac{\rho + p}{3H} \Gamma.
\end{equation}
As should be expected, for $\Gamma=0$, the creation pressure and entropy source vanish thereby recovering the  perfect fluid description}.

\vskip 0.4cm

\section{Matter Creation Rate and the Transition Redshift}

In this appendix we show a curious result, namely: the existence of a transition redshift, $z_t$,  
at late time  determines the simplest form of the matter creation rate. 
In order to show that we  consider the evolution equation (see section III)
\begin{equation}
\label{evolRA}
     R\ddot{R} + \frac{1}{2}\left(1 - \frac{\Gamma}{H}\right) \dot{R}^2 = 0,  
\end{equation}
which means that the decelerating parameter ($q=-R {\ddot R}/{\dot R^{2}}$) can be written as:
\beq 
q=\frac{1}{2}\left[1- \frac{\Gamma}{H} \right].
\eeq
The above expression was first obtained by Zimdahl et al. \cite{ZSBP01} using a different notation (see their Eq. (53)).  

Now, by taking $q(z_t)=0$  in the above expression one finds that $\Gamma = H (z_t)$, 
the value of the Hubble parameter at the instant of transition. At low redshifts it 
is natural to take it proportional to $H_0$, say, $\Gamma= 3\gamma H_0$, where the factor 3 
is introduced for mathematical convenience and the  constant $\gamma$ parameter, in general, depends 
on the transition redshift (see Eq. (\ref{gammazt})). 
Note also that the $\beta$ contribution can be thought as the first 
order correction of this quantity in powers of $H/H_0$
\begin{equation} 
\frac{\Gamma}{3\gamma H_0}= 1 + \frac{\beta}{\gamma}{\frac{H}{H_0}} +...,
\end{equation}
and, therefore, we may write 
\begin{equation} 
q = \frac{1}{2}\left[1 - 3\beta - 3\gamma \frac{H_0}{H} \right],
\end{equation}
which is the same expression appearing in section IV (see Eq. (\ref{q1})). 
For $\gamma=0$, the resulting scenario was proposed by Lima, Germano and Abramo \cite{LGA96} (see also Refs. \cite{LA99}) while for $\beta=0$, it was first discussed by Zimdahl et al. \cite{ZSBP01}. Clearly, the scenario proposed here is a combination of both approaches. Note also that only in the enlarged form, it may represent a possible solution to the old (and modified versions) of the coincidence problem (see the available space parameter in the ($\gamma,\beta$) plane as shown in Fig. 4b).  

\begin{acknowledgments}
The authors would like to thank V. Busti, J. V. Cunha,  A. C. C. Guimar\~aes, R. Holanda, J. F.
Jesus and L. Sodr\'e for helpful discussions. JASL is partially supported by
CNPq and FAPESP under Grants 304792/2003-9 and 04/13668-0, respectively. 
RCS is supported by CNPq No. 15.0293/2007-0 and EFS by CAPES (Brazilian
Research Agencies).
\end{acknowledgments}

\end{document}